\documentclass[options,onecolumn]{mn2e}
\usepackage{psfig}

\newif\ifAMStwofonts

\title[Primordial non-Gaussianity] 
{Primordial non-Gaussianity: local
curvature method and statistical significance of constraints on 
$f_{\rm NL}$ from WMAP data}
\author[P. Cabella, M. Liguori, F.~K. Hansen, D. Marinucci, S. Matarrese, 
L. Moscardini, N. Vittorio]
{P. Cabella,$^1$\thanks{E-mail: paolo.cabella@roma2.infn.it}
M. Liguori,$^2$\thanks{E-mail:michele.liguori@pd.infn.it} F.~K. Hansen,$^1$\thanks{Frode.Hansen@roma2.infn.it} D. Marinucci,$^3$\thanks{marinucc@mat.uniroma2.it} S. Matarrese,$^2$\thanks{sabino.matarrese@pd.infn.it}
\newauthor L. Moscardini,$^4$\thanks{lauro.moscardini@unibo.it} N. Vittorio$^{1,5}$\thanks{vittorio@roma2.infn.it}\\
$1$ Dipartimento di Fisica, Universit\`a di Roma `Tor Vergata', Via
della Ricerca Scientifica 1, I-00133 Roma, Italy\\
$^2$  Dipartimento di Fisica 'Galileo Galilei', 
Universit\`a di Padova and INFN, Via Marzolo 8, I-35131 Padova, Italy\\
$^3$ Dipartimento di Matematica, Universit\`a di Roma `Tor Vergata', Via
della Ricerca Scientifica 1, I-00133 Roma, Italy  \\        
$^4$ Dipartimento di Astronomia, Universit\`a di Bologna , 
Via Ranzani 1,I-40127 Bologna, Italy\\
$5$ INFN, Sezione di Roma `Tor Vergata', Via
della Ricerca Scientifica 1, I-00133 Roma, Italy\\}

\pagerange{\pageref{firstpage}--\pageref{lastpage}}
\pubyear{2004}

\begin{document}

\label{firstpage}

\maketitle

\begin{abstract}
We test the consistency of estimates of the non-linear coupling
constant $f_{\rm NL}$ using non-Gaussian CMB maps generated by the method described in (Liguori, Matarrese and Moscardini 2003). This procedure to obtain non-Gaussian maps differs significantly from the
method used in previous works on estimation of $f_{\rm NL}$. Nevertheless, using
spherical wavelets, we find results in very good agreement with
(Mukherjee and Wang 2004), showing that the two ways of generating primordial 
non-Gaussian maps give equivalent results. Moreover, we introduce a
new method for estimating the non-linear coupling constant from CMB
observations by using the local curvature of the temperature
fluctuation field.   We present both Bayesian credible regions
(assuming a flat prior) and proper (frequentist) confidence intervals
on $f_{\rm NL}$, and discuss the relation between the two
approaches. The Bayesian approach tends to yield lower error bars than the frequentist approach, suggesting that a careful analysis of the different interpretations is needed. Using this method, we estimate $f_{\rm NL}=-10^{+270}_{-260}$ at the $2\sigma$ level (Bayesian) and $f_{\rm NL}=-10^{+310}_{-270}$ (frequentist). Moreover, we find that the wavelet and the local curvature approaches, which provide similar error bars, yield approximately uncorrelated estimates of $f_{NL}$ and therefore, as advocated in (Cabella et al. 2004), the estimates may be combined to reduce the error bars. In this way, we obtain $f_{NL}=-5\pm 85$ and $f_{NL}=-5\pm 175$ at the $1\sigma$ and $2\sigma$ level respectively using the frequentist approach.
\end{abstract}


\section{Introduction}

Inflation is the standard paradigm for providing the initial
conditions for structure formation and Cosmic Microwave Background (CMB)
anisotropy generation. In
the inflationary picture, primordial adiabatic perturbations arise
from quantum fluctuations of the {\it inflaton} scalar field which
drives the accelerated Universe expansion. In the simplest
models, the inflaton is assumed to have a shallow potential, thereby
leading to a slow rolling of this field down its potential. The
flatness of the potential implies that intrinsic non-linear (hence
non-Gaussian) effects during slow-roll inflation are tiny, although
non-zero and calculable \cite{frs,gan,lps,wk,acqua,malda}. To
quantitatively describe the theoretical findings in this framework,
let us introduce a useful parameterisation of non-Gaussianity
according to which the primordial gravitational potential $\Phi$ is
given by a linear Gaussian term $\phi_{\rm G}$, plus a quadratic contribution,
as follows (e.g. \cite{verde}):
\begin{equation}
\label{eq:model}
\Phi({\bf x}) = \phi_{\rm G}({\bf x}) + f_{\rm NL} \phi_{\rm G}^2({\bf x})
\end{equation}
(up to a constant offset, which only affects the monopole
contribution), where the dimensionless parameter $f_{\rm NL}$ sets the
strength of non-Gaussianity. The above mentioned calculation
of the amount of non-Gaussianity during single-field inflation leads
to typical values $f_{\rm NL} \sim 10^{-2}$, much too low to be
observable in CMB experiments. However, non-linear gravitational
corrections after inflation unavoidably and significantly enhance the
non-Gaussianity level, leading to values of $f_{\rm NL} \sim 1$,
almost independent of the detailed inflation dynamics
\cite{bmr04a}. An angular modulation of the quadratic 
term is also found \cite{bmr04b}, 
so that $f_{\rm NL}$ should be considered as a
kernel in Fourier space, rather than a constant.  The resulting
effects in harmonic space might be used to search for signatures of
inflationary non-Gaussianity in the CMB \cite{michele2}.  Nonetheless,
owing to the large values of $|f_{\rm NL}|$ considered here ($\sim 10^2$)
we will disregard this complication and assume $f_{\rm NL}$ to
be a constant parameter.  Despite the simplicity of the inflationary
paradigm, the mechanism by which adiabatic (curvature) perturbations
are generated is not yet fully established. In the {\it standard
scenario} associated to single-field models of inflation, the observed
density perturbations are due to fluctuations of the inflaton field,
driving the accelerated expansion.  An alternative to the standard
scenario which has recently gained increasing attention is the {\it
curvaton} mechanism \cite{enq,mt,mt2,lyth,lythwands,bmr04c}, 
according to which the final curvature 
perturbations are produced from an initial isocurvature perturbation
associated to the quantum fluctuations of a ``light'' scalar field
other than the inflaton, the so-called ``curvaton'', whose energy
density is negligible during inflation. Due to a non-adiabatic
pressure perturbation arising in multi-fluid systems \cite{mol90}
curvaton isocurvature perturbations are transformed into adiabatic
ones, when the curvaton decays into radiation much after the end of
inflation.  Another recently proposed mechanism for the generation of
cosmological perturbations is the {\it inhomogeneous reheating}
scenario \cite{dvali,zalda,kof03,mr}.
It acts during the reheating stage after inflation if super-horizon
spatial fluctuations in the decay rate of the inflaton field are
induced during inflation, causing adiabatic perturbations in the final
reheating temperature in different regions of the universe.  An
important feature of both the curvaton and inhomogeneous reheating
scenarios is that, contrary to the single-field slow-roll models, they
may naturally lead to high levels of non-Gaussianity.  Large levels of
non-Gaussianity are also predicted in a number of theoretical variants
of the simplest inflationary models.  First, generalised multi-field
models can be constructed in which the final density perturbation is
either strongly \cite{sal90,sal91,kof03} or mildly
\cite{bmr02,bernaa,bernaa2,enq04} non-Gaussian, and
generally characterised by a cross-correlated mixture of adiabatic and
isocurvature perturbation modes \cite{wands}. Values of $|f_{\rm NL}|
\sim 10^2$ are also predicted in the recently proposed {\it
ghost-inflation} picture \cite{arkani}, as well as in theories based
on a Dirac-Born-Infeld (DBI)-type Lagrangian for the inflaton
\cite{ast}.  Quite recently, there has been a burst of interest for
non-Gaussian perturbations of the type of Eq.~(\ref{eq:model}).
Different CMB datasets have been analysed, with a variety of
statistical techniques
(e.g. \cite{komatsu,cayon,santos,wang,gaztanaga}) with the aim of
constraining $f_{\rm NL}$.  In the last years some authors set
increasingly stringent limits on the primordial non-Gaussianity level
in the CMB fluctuations. Using a bispectrum analysis on the COBE DMR
data \cite{komatsua} found $|f_{\rm NL}|< 1500$. On the same data,
\cite{cayona} found $|f_{\rm NL}|< 1100$ using Spherical Mexican Hat
Wavelets (SMHW) and \cite{santos} using the MAXIMA data set the limit
on primordial non-Gaussianity to be $|f_{\rm NL}|< 950$. All these limits
are at the 1$\sigma$ confidence level. The most stringent limit to
date has been obtained by the {\em WMAP} team \cite{komatsu}: $-58
<f_{\rm NL}< 134$ at $95~\%$ cl. Consistent results (an upper limit of
$f_{\rm NL}\le220$ at a $2\sigma$ confidence level) have been obtained
from the WMAP data using SMHW \cite{wang}. It was shown in \cite{ks}
that the minimum value of $|f_{\rm NL}|$ which can be in principle
detected using the angular bispectrum, is around 20 for {\em WMAP}, 5
for {\em Planck} and 3 for an {\em ideal} experiment, owing to the
intrinsic limitations caused by cosmic variance.  Alternative
strategies, based on the multivariate empirical distribution function
of the spherical harmonics of a CMB map, have also been proposed
\cite{hanmarvit,hansenetal}, or measuring the trispectrum of the CMB
\cite{grazia}.

The plan of the paper is as follows: in Section \ref{sect:mapng} we
describe our method to produce the temperature pattern of the CMB in
presence of primordial non-Gaussianity; Section \ref{sect:test}
addresses statistical issues to constrain the non-linearity parameter
$f_{\rm NL}$ on the basis of WMAP data; finally, in Section
\ref{sect:concl} we draw our conclusions.
   
\section{Map making of primordial Non-Gaussianity} 
\label{sect:mapng} 

The non-Gaussian CMB maps used in the following analysis have been
generated by applying the numerical algorithm introduced by
\cite{michele}, hereafter LMM.  Here we summarise the various steps
that define the whole procedure and refer the reader to that paper for
further details.
 
The starting point of the method is the simulation, directly in real
space, of independent complex Gaussian variables $n_{\ell m}(r)$,
with correlation functions
\begin{equation} 
\label{eqn:whitenoise} 
\left \langle n_{\ell_1 m_1}(r_1)
n^*_{\ell_2 m_2}(r_2) \right \rangle = 
\frac{\delta^D(r_1-r_2)}{r^2}\delta_{\ell_1}^{\ell_2} \delta_{m_1}^{m_2}\; ,  
\end{equation}
where $\delta^D$ is the Dirac delta function and $\delta^i_j$ is
Kronecker's delta.

The linear potential multipoles $\Phi^{\rm L}_{\ell m}(r)$ 
(here $\Phi^{\rm L} \equiv \phi_{\rm G}$) having the
desired correlation properties, as expressed in terms of the
primordial (i.e. unprocessed by the radiation transfer function)
power-spectrum of the gravitational potential $P_\Phi(k)$, can be
obtained by convolving $n_{\ell m}(r)$ with suitable filter functions
$W_\ell(r,r_1)$:
\begin{equation}
\label{eqn:real_convolution} 
\Phi^{\rm L}_{\ell m}(r) = \int \! dr_1 \, r_1^2 \, n_{\ell m}(r_1) 
W_\ell(r,r_1) \; .
\end{equation} 
The filters $W_\ell(r,r_1)$ are defined as (see LMM)
\begin{equation}
\label{eqn:filter} 
W_\ell(r,r_1) =
\frac{2}{\pi} \int \! dk \, k^2 \, \sqrt{P_\Phi(k)} \, j_\ell(kr)
j_\ell(kr_1) \; ; 
\end{equation} 
here $j_\ell$ are spherical Bessel function of order $\ell$.  Notice
that $W_\ell(r,r_1)$ can be pre-computed at the beginning of all
simulations and then applied in Eq.~(\ref{eqn:filter}).  One more
advantage of this approach is that $W_\ell(r,r_1)$, at fixed $r$, are
smooth functions of $r_1$ which differ from zero only in a narrow
region around $r_1=r$; as a consequence, the integral in
Eq.~(\ref{eqn:real_convolution}) can be estimated in a fast way by
computing $W_\ell(r,r_1)$ in few points.

At this point the values of the linear potential $\Phi^{\rm L}({\bf
r})$ can be recovered thanks to its expansion in spherical harmonics:
\begin{equation} 
\label{eqn:phi_lin} 
\Phi^{\rm L}({\bf r}) = \sum_{\ell m} \Phi^{\rm L}_{\ell m}(r) Y_{\ell m}
({\hat r})\ .
\end{equation}
The non-Gaussian contribution (modulo $f_{\rm NL}$) to the
gravitational potential, $\Phi^{\rm NL}({\bf r})$ (here 
$\Phi^{\rm NL} \equiv \phi_{\rm G}^2$), is obtained
directly in spherical coordinates by squaring $\Phi^{\rm L}$; this is
then harmonic-transformed by using the HEALPIX package \cite{healpix}
to get $\Phi^{\rm NL}_{\ell m}(r)$.

Finally, the linear and non-linear contributions to the total CMB
multipoles $a_{\ell m} \equiv a_{\ell m}^{\rm L} + f_{\rm NL} a_{\ell
m}^{\rm NL}$ are obtained by convolving each term with the real-space
radiation transfer function $\Delta_\ell(r)$,
\begin{equation}
\label{eqn:rtf_definition} 
\Delta_{\ell}(r) \equiv
\frac{2}{\pi} \int \! dk \, k^2 \, \Delta_\ell(k) j_\ell(kr) \; 
\end{equation}
(see LMM, for the formal derivation),
\begin{eqnarray}
\label{eqn:rtf}
a_{\ell m}^{\rm L} & = & \int \! dr \, r^2 \Phi_{\ell m}^{\rm L}(r) 
\Delta_{\ell}(r), \nonumber \\ \;
a_{\ell m}^{\rm NL} & = & \int \! dr \, r^2 \Phi_{\ell m}^{\rm NL}(r) 
\Delta_{\ell}(r) \;. 
\end{eqnarray}
Notice that also the quantities $\Delta_\ell(r)$, which have been
numerically estimated by using a modification of the CMBFAST code
\cite{sz} can be pre-computed and stored for all
the simulations of a given model.

In \cite{komatsu,komat2} a different approach to produce
non-Gaussian maps was adopted. Their starting point is the generation on a
Fourier-space grid of a Gaussian field, which is then inverse-Fourier 
transformed and squared to get the non-Gaussian part of the
gravitational potential in real space. The successive steps involve
interpolation on a spherical grid, harmonic transforms and convolution
with $\Delta_\ell(r)$, to obtain the Gaussian and non-Gaussian CMB
multipole coefficients.

\section{Tests of Non-Gaussianity} 
\label{sect:test}

In this section, we will estimate $f_{\rm NL}$ from the WMAP data using
two different approaches. The estimation procedures will be calibrated
using non-Gaussian maps produced by the method described above. We
have replicated the method applied in \cite{wang} (hereafter MW) in
order to check that the estimates using non-Gaussian maps generated
with two different methods, give consistent results. We also introduce
a new estimator of $f_{\rm NL}$ based on the local curvature properties of
the CMB fluctuation field. Finally, we observe that the estimates of $f_{\rm NL}$ from the two methods are approximately uncorrelated; therefore we introduce a combined estimator which reduces the error bars by a factor of about $\sqrt{2}$.

\subsection{Local curvature}
The local curvature test has been used in the flat limit to test the
presence of non-Gaussianity due to cosmic strings \cite{dore}, and the
extension to the spherical case \cite{frode,paolo}, has been used to
verify the asymmetries in the WMAP data. In this section we will test
the power of the local curvature test to detect primordial
non-Gaussianity. First we review the method.\\ We consider a CMB
temperature map $T(\theta,\phi)$ normalised with its standard
deviation $\sigma$:
\begin{equation}
T(\theta,\phi)\rightarrow \frac{T(\theta,\phi)-\bar{T}}{\sigma}.
\end{equation}    
The Hessian of this map can be written as 
\begin{equation}
\label{eq:hess}
H=\left(
\begin{array}{cc}
T_{;\theta\theta} & T_{;\theta\phi}\\
T_{;\phi\theta} & T_{;\phi\phi} \end{array} \right)
=
\left(
\begin{array}{cc}
T_{,\theta\theta} & (T_{,\theta\phi}-\cot\theta T_{,\phi})/\sin\theta \\
(T_{,\theta\phi}-\cot\theta T_{,\phi})/\sin\theta  &
(T_{\phi\phi}+1/2\sin2\theta T_{,\theta})/\sin^2\theta \\
\end{array}
\right) , 
\end{equation}
where the comma denotes the ordinary derivative and the semicolon covariant
derivative.
In order to evaluate the derivatives it is convenient to go to
harmonic-space:
\begin{equation}
T(\theta,\phi)_{,i}=\sum a_{lm}Y_{lm}(\theta,\phi)_{,i}
\end{equation}
and to use the recurrence relations \cite{vals}:
\begin{eqnarray}
\label{eq:vals}
\frac{\partial}{\partial\phi}Y_{lm}(\theta,\phi)&=&im Y_{lm}(\theta,\phi)\ ,\\
\frac{\partial}{\partial\theta}Y_{lm}(\theta,\phi)&=&\frac{1}{2}\sqrt{l(l+1)-
m(m
+1)}Y_{lm+1}(\theta,\phi)e^{-i\phi}-\frac{1}{2}\sqrt{l(l+1)-m(m-1)}Y_{lm-1}
(\theta,\phi)e^{i\phi},\nonumber
\end{eqnarray}
twice to obtain the Hessian values in every point of the map, as
suggested in \cite{schmalzing}. The presence of noise may render the derivatives unstable, but we have checked that the results in this paper do not change significantly when smoothing the map before performing the derivatives.

The points of the renormalised map
can be classified as:
\begin{itemize}
\item \emph{hills} where the eigenvalues of the Hessian  are both positive,
\item \emph{lakes} where the eigenvalues of the Hessian  are both negative,
\item \emph{saddles} where the eigenvalues have opposite sign.
\end{itemize}
We calculate the proportions of hills, lakes and saddles above a
certain level $\nu$ in the normalised map. In this way, we obtain
three functions of $\nu$ which in the Gaussian case have a known
functional form \cite{dore}. Deviations of these functions from this
Gaussian expectation value can be used to detect non-Gaussianity.

In order to constrain the $f_{\rm NL}$ parameter our procedure is as follows:
\begin{itemize}
\item we generate a set of primordial non-Gaussian maps with the method 
described in Section \ref{sect:mapng} with the WMAP best fit power
spectrum and different $f_{\rm NL}$;
\item we convolve with a beam and add noise corresponding to the 
given experiment;
\item we apply the \emph{Kp0} galaxy and point source mask;
\item we degrade the maps to the resolution of \emph{nside} = 256
where the derivatives are performed;
\item we count the densities of hills, lakes and saddles for each 
map and each level $\nu$ with an extended mask to avoid the
instabilities of the derivatives close to the boundaries of
\emph{Kp0} (for details of the extensions, see \cite{frode}). In this
way, we obtain the form of the hill, lake and saddle densities as a
function of $f_{\rm NL}$;
\item we repeat the last three points for the data of the given 
experiment.
\item for each simulation as well as for the data, we construct a $\chi^2$ as follows
\begin{equation}
\label{eq:chicur}
\chi^2(f_{\rm NL})=(\mathbf{x}-\langle\mathbf{x}\rangle)^T{\mathbf M}^{-1}(\mathbf{x}-\langle\mathbf{x}\rangle),
\end{equation}
where $\mathbf{x}=[h(\nu_0),h(\nu_1),...,h(\nu_{\rm max}),l(\nu_0),l(\nu_1),...,l(\nu_{\rm max})]$, $h(\nu)$ and $l(\nu)$ are the hill and lake densities and $\nu_0,\nu_1,...,\nu_{\rm max}$ are the threshold values used, given by $\nu_0=-3\sigma$ and $\nu_{\rm max}=2\sigma$. The maximum threshold was determined in such a way as to obtain sufficient statistics for the lake proportions which go to zero at high thresholds. The covariance matrix ${\mathbf M}$ with elements $M_{ij}=\langle x_ix_j \rangle$ is evaluated on the basis of 1000 Gaussian simulations. The distribution of the hill and lake densities at each threshold has been found to be close to Gaussian, justifying the above form of the $\chi^2$.
\item The estimate of $f_{\rm NL}$ is obtained for each simulation and for the data by minimising this $\chi^2$ with respect to $f_{\rm NL}$.
\item Bayesian error bars are derived by constructing the likelihood $L(f_{\rm NL})\propto e^{-\chi^2/2}$. Integrating this likelihood with respect to the parameter, we obtain the approximate Bayesian ``credible regions'' (see Section \ref{sect:stati}).
\item The frequentist error bars are derived by the histogram of the estimates $\hat f_{\rm NL}$ from simulations and defining the $1$ and $2\sigma$ levels as the limits within which $68\%$ and $95\%$ of the estimates fall.
\end{itemize}

Note that we will perform the estimates (1) using only the diagonal part of the covariance matrix and (2) including the full covariance matrix. We will later show that this makes a huge difference when considering the error bars, and that care has to be taken when approximating the covariance matrix to be diagonal.

\begin{figure}
\begin{center}
\leavevmode
\psfig {file=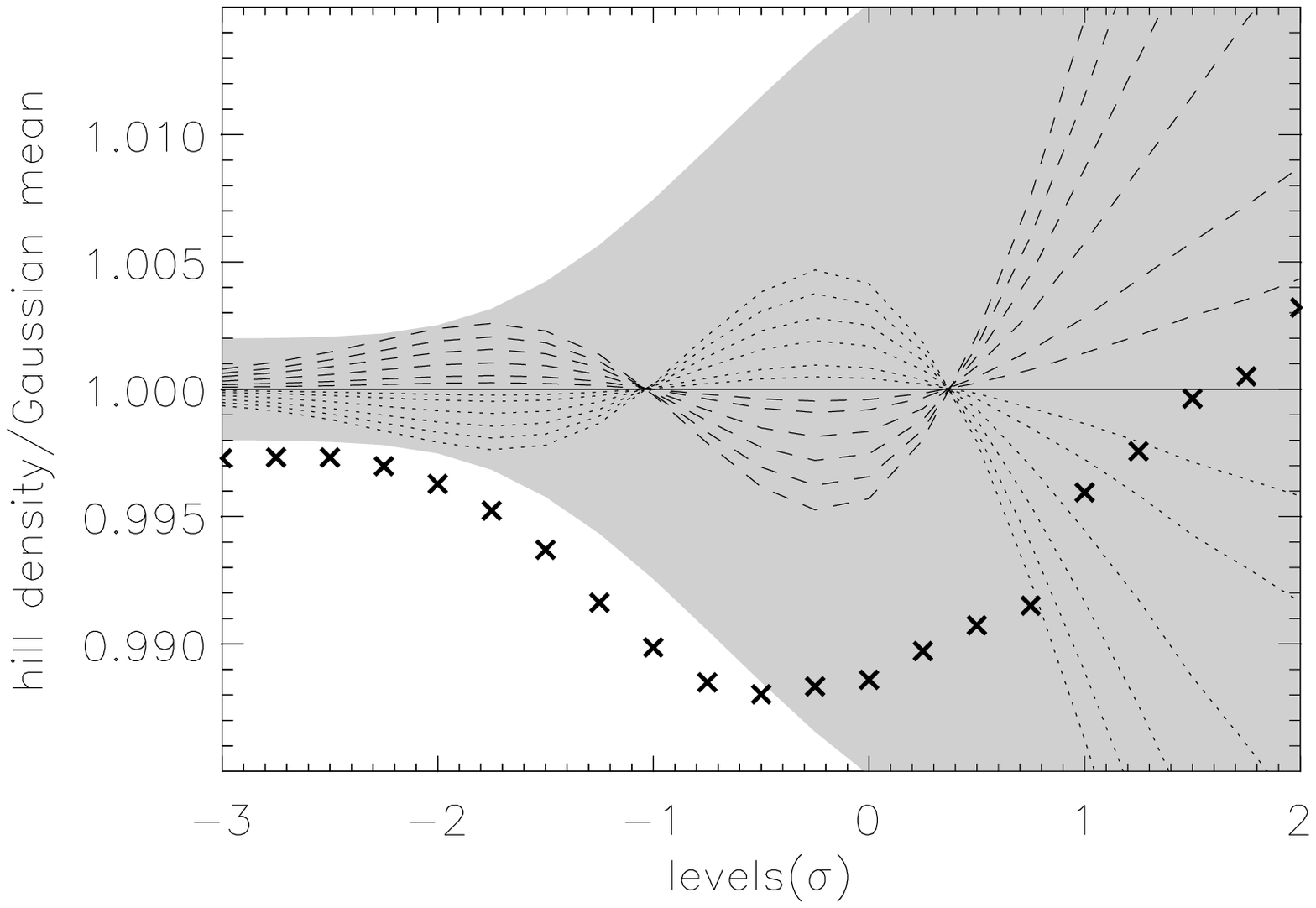,width=8cm,height=8cm}
\psfig {file=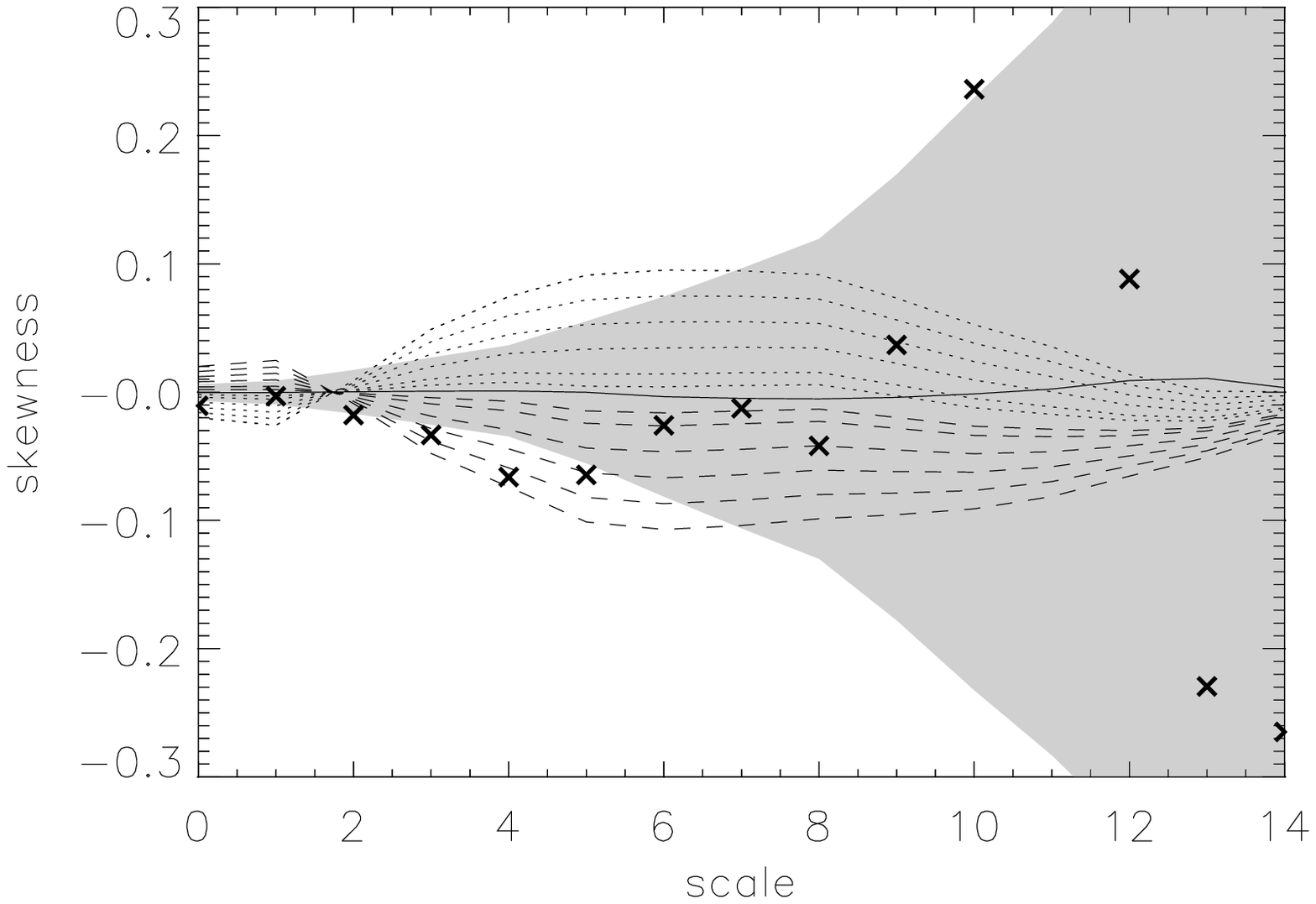,width=8cm,height=8cm}
\caption{On the left, density of hills as a function of the threshold $\nu$, for different 
values of $f_{\rm NL}$. All values are divided by the Gaussian expectation
value. On the right, skewness of wavelet coefficients for different scales $R$. The solid line represents Gaussian simulations, the dotted (dashed)
lines represent non-Gaussian simulations for different negative
(positive) values of $f_{\rm NL}$; the grey band is the $1\sigma$ confidence
level taking into account the noise and beam of the co-added Q+V+W map
with an extended \emph{Kp0} mask. The values of $|f_{\rm NL}|$ shown are
50, 100, 200, 300, 400 and 500, showing an increasing distance from the
Gaussian expectation value. The bold crosses show the results on the
data.}
\label{fig:hillfnl}
\end{center}
\end{figure}

\begin{figure}
\begin{center}
\leavevmode
\psfig{file=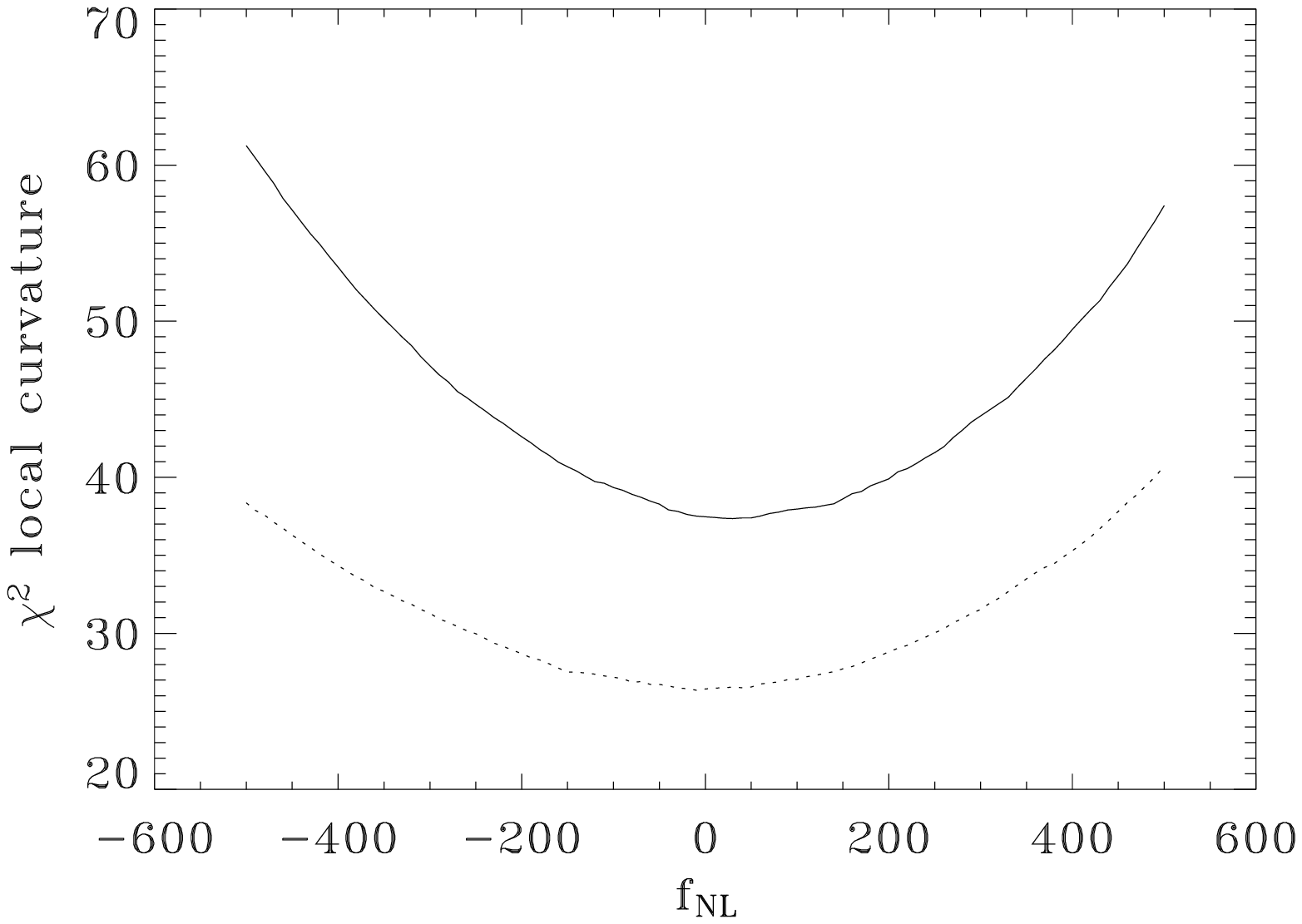,width=8cm,height=8cm}
\psfig {file=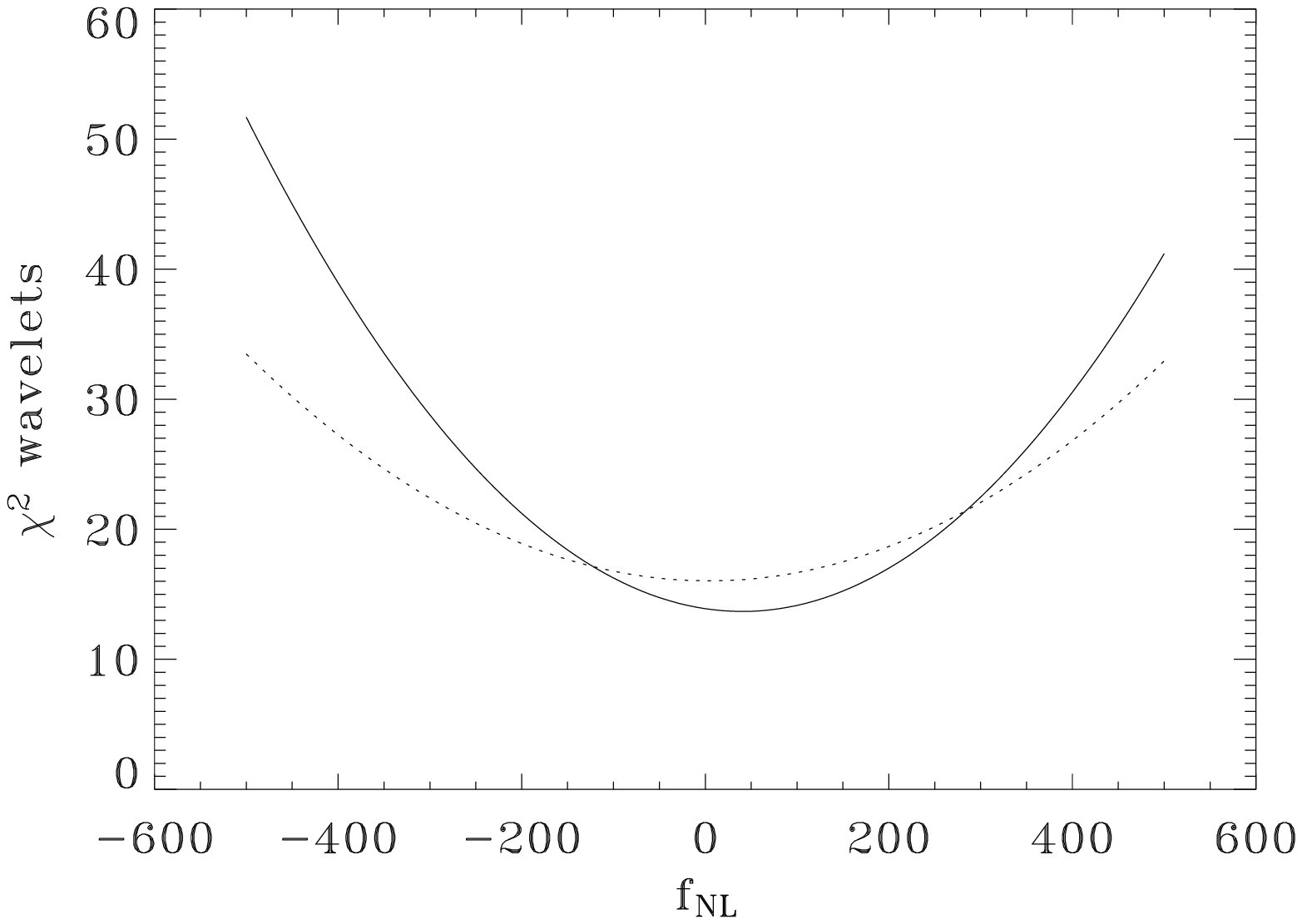,width=8cm,height=8cm}
\caption{The plot shows the $\chi^2$ as a function of $f_{\rm NL}$ around 
its minimum, using the curvature test (left) and the wavelets (right) on the co-added Q+W+V WMAP map. Results including the full covariance matrix are plotted as a dotted line, whereas the solid line corresponds to the diagonal matrix approximation.}
\label{fig:chi2data}
\end{center}
\end{figure}

The results of these simulations are shown in the left panel of Figure~\ref{fig:hillfnl}, where we can see the effect for the different values of
$f_{\rm NL}$ on the hill density. We applied this procedure to the
publicly available WMAP data\footnote{obtainable from the LAMBDA
website: http://lambda.gsfc.nasa.gov/}. We co-added the (foregrounds subtracted) maps from the three WMAP frequency channels Q, U and W following the procedure in \cite{WMAP}. In Figure~\ref{fig:chi2data} we show the $\chi^2({\tiny
f_{\rm NL}})$ for the data around its minimum in the two cases, with and without the off-diagonal elements of the covariance matrix. When using only the diagonal parts of the covariance matrix we estimate $f_{\rm NL} = 30$ with a $68\%$ and $95\%$ credible region  equal to $-50 < f_{\rm NL} < 140$ and $-170 < f_{\rm NL} < 240$, respectively, in good agreement with previously released estimates \cite{komatsu,wang}. We also include proper (frequentist) confidence intervals, which turn out to be $-230 < f_{\rm NL} < 280$ and $-540 <
f_{\rm NL} < 570$, respectively. The relationship between the two
approaches is discussed in subsection \ref{sect:stati}. 
When including the full covariance matrix, we estimate $f_{\rm NL} = -10$ with a $68\%$ and $95\%$ approximate credible region equal to $-120 < f_{\rm NL} < 120$ and $-270 < f_{\rm NL} < 260$ respectively, while the frequentist constraints turn out to be $-140 < f_{\rm NL} < 120 $ and $-280 < f_{\rm NL} < 300$. We observe that when not including the full covariance matrix, there is a huge difference in error bars between the Bayesian/frequentist approaches. Including the full matrix, this difference is smaller but persists. The much bigger frequentist error bars can be explained as follows: when assuming that the thresholds are uncorrelated, huge deviations from the expected value at many thresholds provide strong evidence against the model to be tested. On the other hand, taking into account the fact that the thresholds are highly correlated, the evidence  becomes weaker and the expected model can still be consistent with the observations. Note further that Bayesian error bars tend to be lower than in the frequentist case in the above estimates from the WMAP data; the same phenomenon occurs in simulated maps.\\
As a final
remark, we note that equation (\ref{eq:chicur}) can be exploited to
implement a goodness-of-fit test for our specification of noise and
foreground features. More precisely, we compared the best fit value
$\chi(f_{\rm NL}=30)$ obtained for the WMAP data with 
$\min \chi(f_{\rm NL})$ from
200 Monte Carlo simulations of non-Gaussian CMB maps with
$f_{\rm NL}=30$. When using the diagonal covariance matrix, the observed value corresponds to the $28\%$ quantile, thereby suggesting good agreement between WMAP data and our simulated models. When we adopt the full covariance matrix, the observed value is smaller than $92\%$ of the simulations, that is, the local curvature statistics on WMAP data are closer to their expected values than the great majority of simulated maps. Note however that we are performing four similar tests on the data (local curvature, wavelets with and without covariance matrix) and therefore a $92\%$ deviation from simulations is not significant.

\subsection{Spherical Wavelets}
Wavelets are a very flexible tool used in connection with CMB data 
for denoising \cite{sanz}, extracting point sources \cite{cayon,tenorio}, 
and detecting non-Gaussianity
\cite{cabella,Martinez,barreiro,forni,starck}. To test the consistency
of our maps of primordial non-Gaussian models with those of \cite{komatsu} we
have repeated the method for estimating $f_{\rm NL}$ described in MW. The
basics steps for the test are the following (for more details, see 
MW):
\begin{itemize}
\item we generate a set of primordial non-Gaussian maps with 
the method described in Section \ref{sect:mapng} with the WMAP best
fit power spectrum and different $f_{\rm NL}$;
\item we add noise and convolve with a beam corresponding to the given 
experiment; 
\item we apply the \emph{Kp0} galactic cut, leaving the point sources 
unmasked;
\item we degrade the maps to the resolution of \emph{nside} = 256;
\item we obtain the wavelet coefficients by convolving each map with 
the Spherical Mexican Hat Wavelet (SMHW) given by:
\begin{equation}
\Psi(y,R)=\frac{1}{\sqrt{2\pi}N(R)}
\left[1+\left(\frac{y}{2}\right)^2\right]^2
\left[2-\left(\frac{y}{R}\right)^2e^{-y^2/2R^2}\right],
\end{equation}
using the scales $R$ given in MW (14, 25, 50, 75, 100, 150, 200, 250, 300, 400, 500, 600, 750, 900, 1050 arcmin) ;
\item we use only the coefficients outside an extended \emph{Kp0} mask 
(for details of the
extensions, see \cite{vielva} and MW) to obtain the skewness as a
function of $f_{\rm NL}$. The result is shown in the right panel of Figure~\ref{fig:hillfnl};
\item we finally repeat the last three steps for the data of the 
given experiment.
\item for each simulation as well as for the data, we construct a $\chi^2$ as follows
\begin{equation}
\label{eq:chiwav}
\chi^2(f_{\rm NL})=(\mathbf{x}-\langle\mathbf{x}\rangle)^T{\mathbf M}^{-1}(\mathbf{x}-\langle\mathbf{x}\rangle),
\end{equation}
where the elements of the vector $\mathbf{x}$ are $x_i=S(R_i)$ where $S(R_i)$ is the skewness of the wavelet coefficients on scale $R_i$. The covariance matrix ${\mathbf M}$ with elements $M_{ij}=\langle x_ix_j \rangle$ is evaluated on the basis of 1000 Gaussian simulations. The distribution of the skewness at each scale is close to Gaussian, justifying the above form of the $\chi^2$.
\item For each simulations and for the data we estimate $f_{\rm NL}$ by minimising this $\chi^2$ with respect to the parameter.
\item Bayesian error bars are found by constructing the likelihood $L(f_{\rm NL})\propto e^{-\chi^2/2}$. Integrating this likelihood with respect to the parameter, we obtain the approximate Bayesian credible regions (see Section \ref{sect:stati}).
\item The frequentist error bars are derived by the histogram of the estimates $\hat f_{\rm NL}$ from simulations and defining the $1$ and $2\sigma$ levels as the limits within which $68\%$ and $95\%$ of the estimates fall.
\end{itemize}

Again, the test was performed on the co-added (foreground cleaned) Q+V+W WMAP map. The plot of the $\chi^2$ for the data is shown in Figure~\ref{fig:chi2data}. We estimate the Bayesian credible region, considering only the diagonal part of the covariance matrix (the results presented in MW were obtained using the same approximation), for $f_{\rm NL}$  to be  $40\pm 90$ at the 1$\sigma$ level and $40\pm180$ at the 2$\sigma$ level, in agreement with the values derived in MW (see again the discussion in Section \ref{sect:stati}). Considering the full covariance matrix, we estimate $f_{\rm NL}$  to be  $0\pm 120$ at the 1$\sigma$ level and $0\pm 240$ at the 2$\sigma$ level As for the curvature, we have
also calculated the frequentist confidence intervals which in the first case turn out to be quite larger, $-100<f_{\rm NL}<170$ and $-230<f_{\rm NL}<320$
respectively, while considering the full covariance matrix we find for the error bars on  $f_{\rm NL}$  $-130<f_{\rm NL}< 120$ and $-260<f_{\rm NL}<230$. Again the Bayesian credible regions are generally smaller than the frequentist confidence levels, but considering the full covariance matrix, the difference is reduced. Note further that the Bayesian credible regions are slightly underestimated when approximating the covariance matrix to be diagonal.

As before, we also implemented a goodness-of-fit test
based on Eq.~(\ref{eq:chiwav}); we found that the values on the WMAP data
corresponds to the $32\%$ and $54\%$ quantile obtained from 200 simulations with $f_{\rm NL}= 40$ when using the diagonal and full covariance matrices, respectively. Again, this suggests that our specification of noise and
foreground masks provides a reasonable approximation to the 
experimental settings of WMAP.


\subsection{The combined test}
In this section we apply the combined procedure introduced in \cite{cabella} to improve the constraints on $f_{\rm NL}$. First of all, we estimated the correlation between the two estimators $\hat{f}_{\rm NL}^{wav}$ and $\hat{f}_{\rm NL}^{cur}$ by 200 Monte Carlo simulations and found
\begin{equation}
\frac{\langle\hat{f_{\rm NL}}^{cur}\hat{f_{\rm NL}}^{wav}\rangle-\langle\hat{f_{\rm NL}}^{cur}\rangle\langle\hat{f_{\rm NL}}^{wav}\rangle}{\sigma^{cur}\sigma^{wav}}\simeq 0.086,
\end{equation} 
suggesting that the correlation between the estimators is less than $10\%$.
Then, we can lower the error bars on $f_{\rm NL}$ by combining the two statistics. More precisely, we use the estimates obtained with the full covariance matrix and we evaluate the frequentist confidence intervals on the following statistic:
\begin{equation}
\hat{f}_{\rm NL}^{comb}=\frac{\hat{f}_{\rm NL}^{wav}+\hat{f}_{\rm NL}^{cur}}{2}.
\end{equation}
Using the combined test on the WMAP data, we estimate  $f_{\rm NL}=-5$,  with the constrains at $1\sigma$ and $2\sigma$ levels of $-90<f_{\rm NL}< 80 $ and $-180 <f_{\rm NL}<170$, respectively.

\subsection{A remark on confidence intervals and credible regions on 
$f_{\rm NL}$}
\label{sect:stati}
In this subsection we present a brief discussion on the evaluation of
(frequentist) confidence intervals and (Bayesian) credible regions on
$f_{\rm NL}$. Let us denote by $\widehat{f}_{\rm NL}^{\ast}$ our estimate of
the non-linearity parameter obtained by minimising
Eqs.~(\ref{eq:chicur}) and (\ref{eq:chiwav}): also, let us denote by
$L(f_{\rm NL};\widehat{f}_{\rm NL}^{\ast})$ the likelihood function. As
well known, an $(1-\alpha)$-level confidence interval for $f_{\rm NL}$
based upon the observation $\widehat{f}_{\rm NL}^{\ast}$ is the set of
all values $f_{\rm NL}$ such that
\[
\int_{|\widehat{f}_{\rm NL}-f_{\rm NL}|\ge|\widehat{f}_{\rm NL}^{\ast}-
f_{\rm NL}|}
L(f_{\rm NL};\widehat{f}_{\rm NL})d\widehat{f}_{\rm NL}\ge\alpha .
\]
Note that the integral is taken with respect to the estimate 
$\widehat{f}_{\rm NL}$ that is, we are viewing $L(f_{\rm NL};
\widehat{f}_{\rm NL})$ as the probability density of
our estimator. In words, we include in the confidence interval all the values
$f_{\rm NL}$ such that the probability to get an estimated parameter as
$\widehat{f}_{\rm NL}^{\ast}$ or further away is at least as large as $\alpha$.
More clearly, a value is included provided it does not entail that
{\it observing what we observed} is less probable than $\alpha$. Of course, 
in the special case where the distribution of $\widehat{f}_{\rm NL}$ is
Gaussian with mean $f_{\rm NL}$ and variance $\sigma^{2}$ which does not
depend on $f_{\rm NL}$ (or, in general, is symmetric under the exchange of
$f_{\rm NL}$ and $\widehat f_{\rm NL}$), we have
\begin{eqnarray}
&&  \int_{|\widehat{f}_{\rm NL}-f_{\rm NL}|\ge|
\widehat{f}_{\rm NL}^{\ast}-f_{\rm NL}
|}L(f_{\rm NL};\widehat{f}_{\rm NL})d\widehat{f}_{\rm NL}\\
&&  =\int_{|\widehat{f}_{\rm NL}-f_{\rm NL}|\ge|
\widehat{f}_{\rm NL}^{\ast}-f_{\rm NL}|}
\frac{1}{\sqrt{2\pi}\sigma}\exp\left\{  -\frac{1}{2}(\widehat{f}_{\rm NL}
-f_{\rm NL})^{2}\right\}  d\widehat{f}_{\rm NL}\\
&&  =\int_{|\widehat{f}_{\rm NL}-f_{\rm NL}|\ge|
\widehat{f}_{\rm NL}^{\ast}-f_{\rm NL}|}
\frac{1}{\sqrt{2\pi}\sigma}\exp\left\{  -\frac{1}{2}(\widehat{f}_{\rm NL}
-f_{\rm NL})^{2}\right\}  df_{\rm NL}
\end{eqnarray}
by the symmetry of the previous expression with respect to
$(\widehat{f}_{\rm NL},f_{\rm NL})$. This justifies the common practice to
derive confidence intervals by integrating the likelihood. Rigorously speaking, however, this is no longer justified
if the integrand is not symmetric with respect to an exchange
of $f_{\rm NL}$ with $\widehat{f}_{\rm NL}$ (for instance if $\sigma$ is not
constant with respect to $f_{\rm NL}$). One may then try to justify the
integration of the likelihood by a Bayesian viewpoint, by assuming a
flat prior and viewing $L(f_{\rm NL};\widehat{f}_{\rm NL})$ as a posterior
density function. The resulting set, however, should not be labelled as
confidence interval (which is a frequentist concept): it is a Bayesian
credible region, which will depend in general on the choice of the
prior.

It is occasionally stated that this dependence is overcome by the
choice of a flat prior. The latter is claimed to be non-informative by
definition: indeed, no physicist would consider a priori equally
likely that $f_{\rm NL}$ lies in $[-10,10]$ rather than it is in
$[317,337]$ (say); thus a flat prior, although unphysical, is usually
justified as a {\it panacea} to get objective results. This argument is to
some extent misleading, though, as it can be shown by standard
counterexamples. Take for instance Eq.~(\ref{eq:model}), and assume
for brevity's sake that $f_{\rm NL}>0$ (otherwise duplicate our argument
by symmetry). Then Eq.~(\ref{eq:model}) can be rewritten as
\begin{equation}
T=\phi_{\rm G}({\bf x})+(g_{\rm NL} \phi_{\rm G}({\bf x}))^{2}\;, 
\label{1bis}
\end{equation}
that is $f_{\rm NL}=g_{\rm NL}^{2}$. Of course, from the physical point of
view there is no reason to prefer the alternative specification in
Eq.~(\ref{eq:model}) to (\ref{1bis}) . Now let us assume we impose a
``non-informative'' (flat) prior on $g_{\rm NL}:$, the posterior
probability becomes
\[
\frac{1}{\sqrt{2\pi}\sigma}\exp\left\{  -\frac{1}{2}(\widehat{f}_{\rm NL}
-g_{\rm NL}^{2})^{2}\right\}  
dg_{\rm NL}=\frac{1}{\sqrt{2\pi}\sigma}\exp\left\{
-\frac{1}{2}(\widehat{f}_{\rm NL}-f_{\rm NL})^{2}\right\}  \frac{df_{\rm NL}}
{2\sqrt{f_{\rm NL}}}
\]
and the credible sets are thus obviously affected, although we are
working with exactly the same model as before and we are claiming to
have used no a priori information. In short, flat priors are simply
shifting the choice from the form of the prior probability to the form
of the statistical parametrisation; the latter, moreover, is not due
to physical considerations, but simply to computational convenience.

\begin{figure}
\begin{center}
\leavevmode
\psfig {file=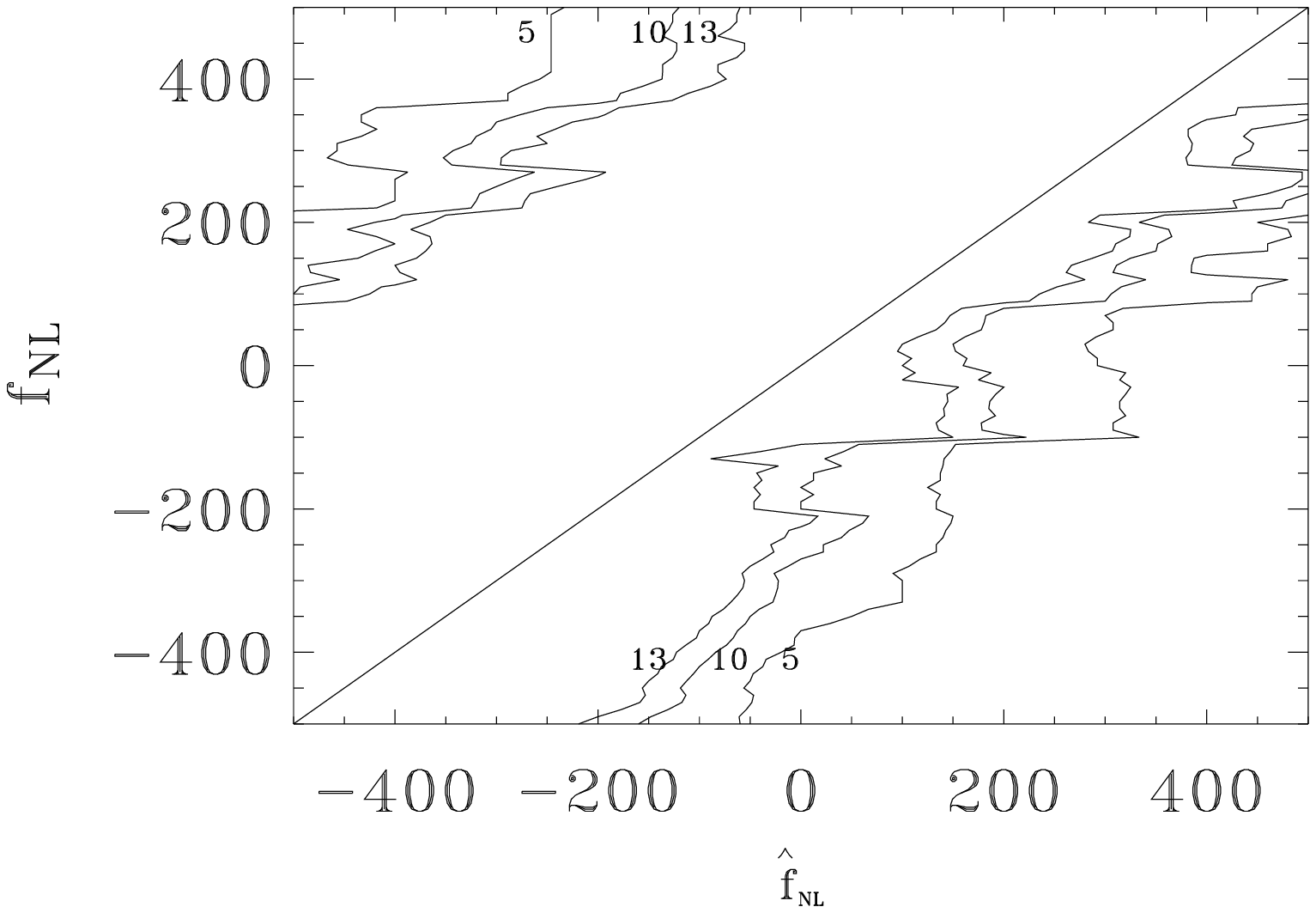,width=8cm,height=8cm}
\psfig {file=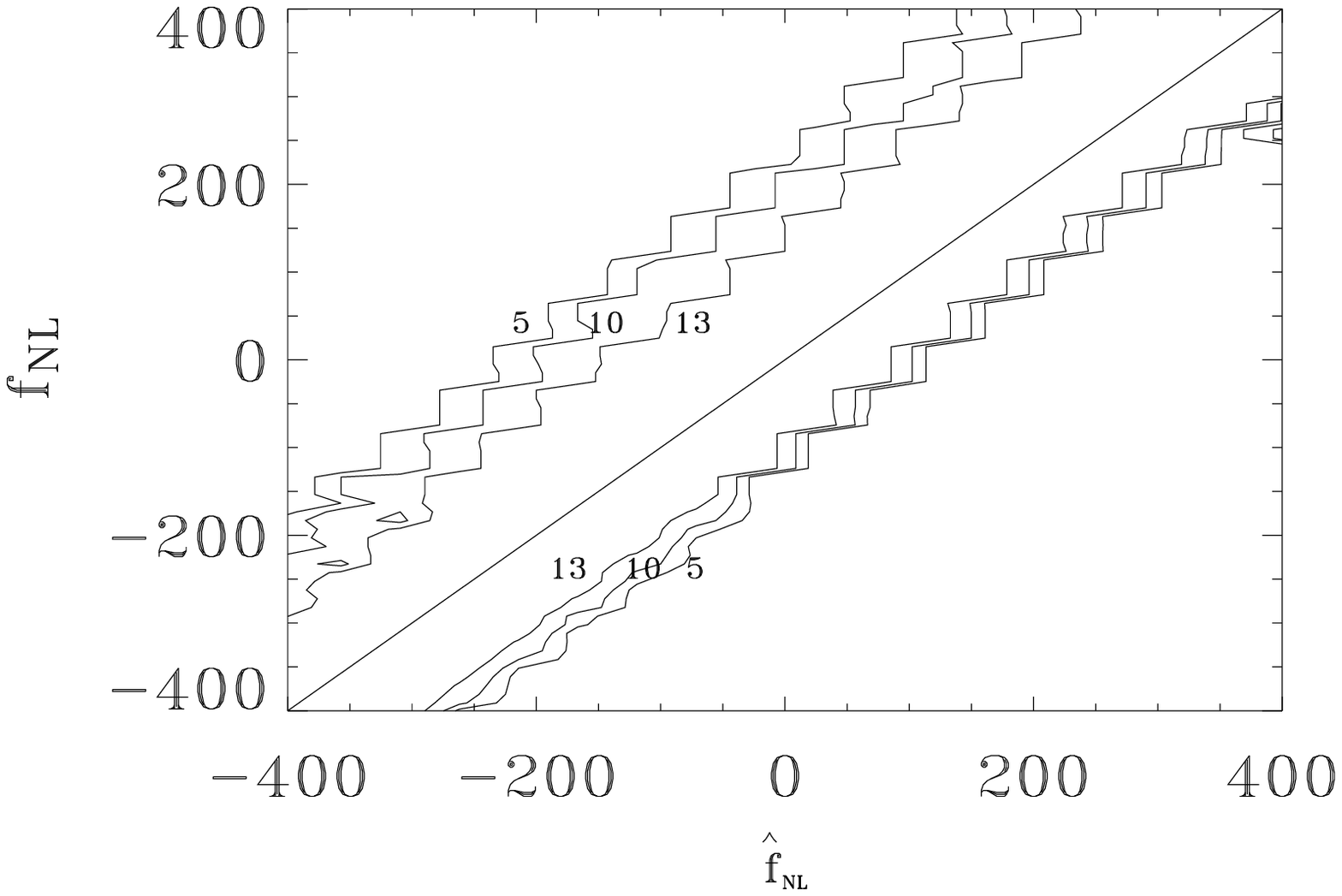,width=8cm,height=8cm}
\caption{The distribution of estimates $\widehat f_{NL}$ as a function of the model parameter $f_{NL}$. The isocontours of the distribution is plotted in terms of percentages of the number of maps used in the simulations. The distribution is clearly skewed with respect to the diagonal shown for clarity. The estimates of $f_\textrm{NL}$ were obtained using the curvature method (left plot) and wavelets (right plot).}
\label{fig:contcur}
\end{center}
\end{figure}

\begin{figure}
\begin{center}
\leavevmode
\psfig {file=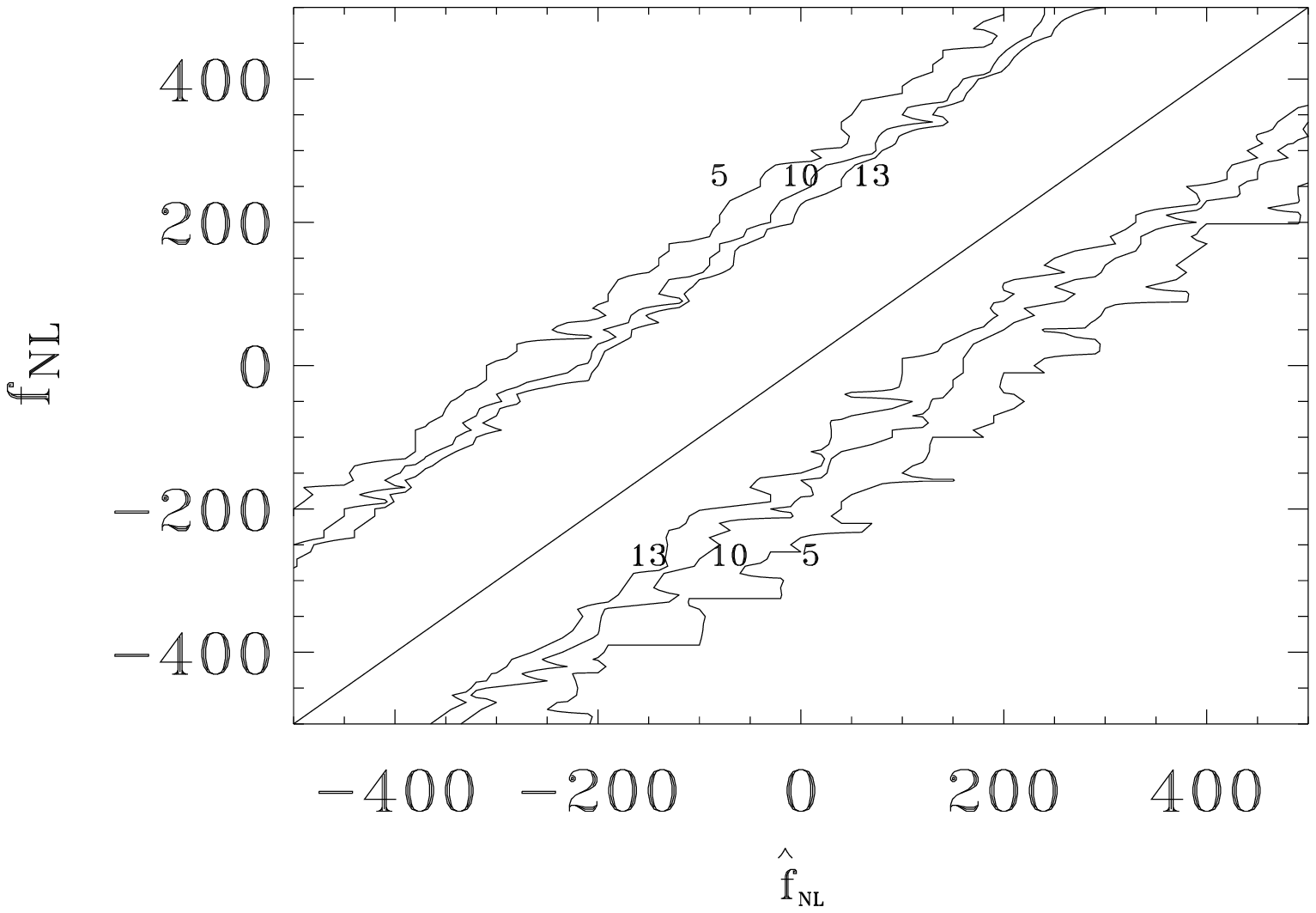,width=8cm,height=8cm}
\psfig {file=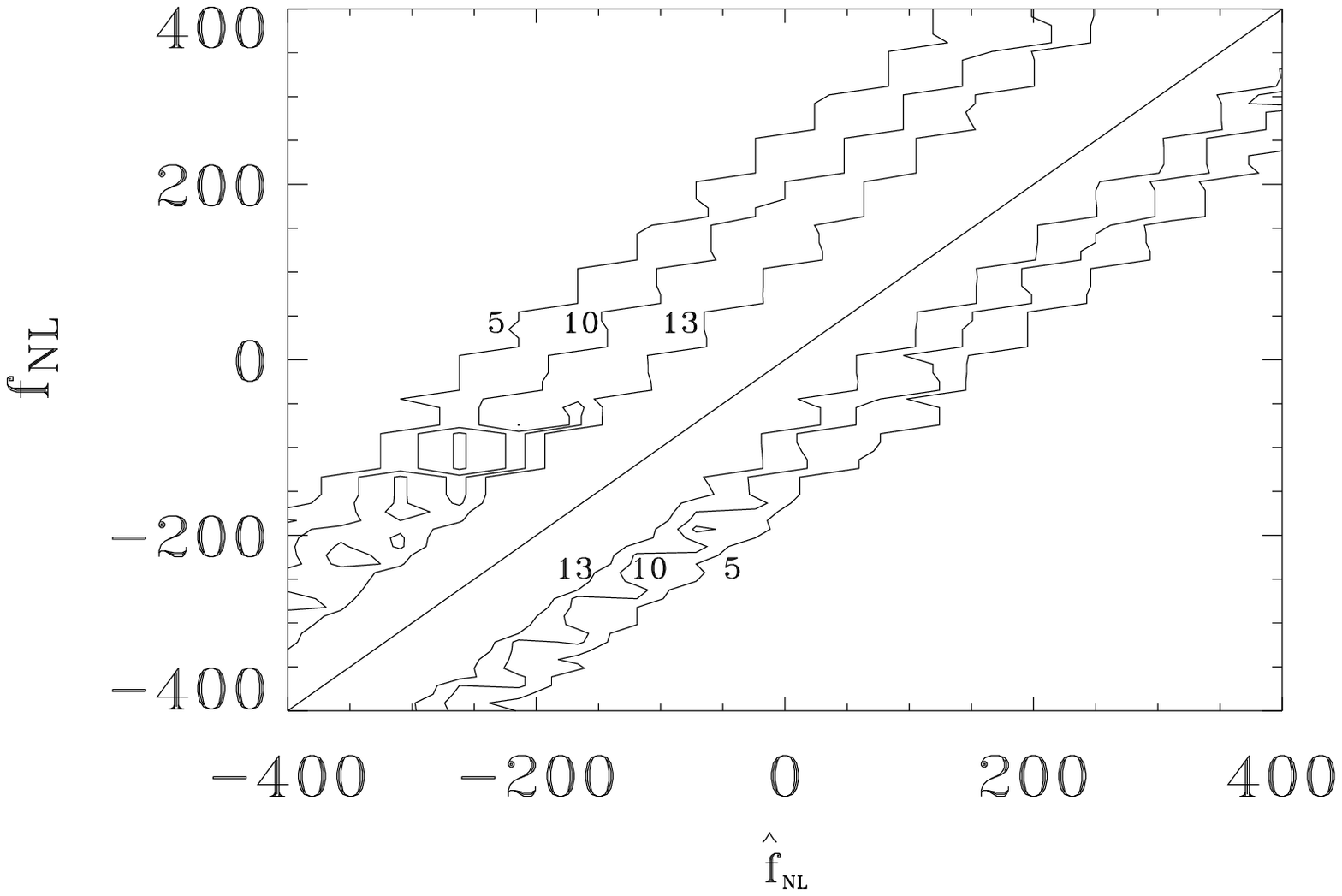,width=8cm,height=8cm}
\caption{As Fig. \ref{fig:contcur} but by using the full correlation matrix. The skewness of the distribution is clearly reduced with respect to the diagonal case (see Fig. \ref{fig:contcur}). The estimates of $f_\textrm{NL}$ were obtained using the curvature method (left plot) and wavelets (right plot).}
\label{fig:contcurcov}
\end{center}
\end{figure}

With this in mind, we will now use $\widehat f_{\rm NL}$ estimated from
200 simulations (with the experimental settings used above) to
investigate whether the distribution of $\widehat f_{\rm NL}$ as a
function of the model $f_{\rm NL}$ is symmetric with respect to an
exchange of $f_{\rm NL}$ and $\widehat f_{\rm NL}$. In 
Figures~(\ref{fig:contcur}), (\ref{fig:contcurcov}) we show a contour plot of
this function for the curvature and wavelet test, 
respectively. When not considering the full covariance matrix, this function is markedly skewed for the wavelet test, but even more so for the curvature test (see fig. \ref{fig:contcur}) . This may explain the big difference between the confidence intervals and Bayesian credible regions for the wavelets and the even bigger difference for the
curvature test when correlations are neglected. On the other hand, when correlations are properly taken into account, the model fit is better and the function becomes considerably more symmetric (see fig. \ref{fig:contcurcov}). This is also reflected in the fact that the difference between Bayesian and frequentist error bars is markedly reduced in this case .


\section{Conclusions}
\label{sect:concl}

In this paper we have used simulated CMB maps of primordial non-Gaussian 
models, generated with the algorithm described in \cite{michele} 
to estimate the
non-linear coupling parameter $f_{\rm NL}$ from the WMAP data. As all
other estimates of $f_{\rm NL}$ in the literature have been based on
non-Gaussian CMB maps generated by a different approach
\cite{komatsu}, one of the aims of this paper was to check whether a
different way of generating CMB maps with the same kind of 
primordial non-Gaussianity gives a consistent estimate of $f_{\rm NL}$. 
In order to perform a direct
test of consistency, we applied the estimator of $f_{\rm NL}$ using SMHW
presented in \cite{wang} on the WMAP data. When neglecting the scale-scale correlations, we find that our estimate
of $f_{\rm NL}$ is in full agreement with theirs 
and thus that the two ways of
generating primordial non-Gaussian maps give fully consistent
results. We have also presented a new method to estimate $f_{\rm NL}$
based on the local curvature of the CMB fluctuation field. Our
estimate of $f_{\rm NL}$ with this method on the WMAP data is consistent
with estimates of $f_{\rm NL}$ using other approaches.

Moreover, we point out the importance of including the full covariance matrix in the $\chi^2$ when estimating $f_{\rm NL}$ with these two methods. We also discuss the difference between frequentist confidence intervals and Bayesian credible regions on the estimate of $f_{\rm NL}$. We show that the difference between these two methods for finding error bars can be huge when not including the full covariance matrix, particularly for the curvature estimator where correlations between thresholds are important. For the WMAP data, the Bayesian credible region expands whereas the frequentist confidence intervals shrink when taking into account correlations in the covariance matrix. We find further that the Bayesian credible regions are in general smaller than the frequentist confidence intervals. We conclude that care must be taken when approximating the confidence intervals on $f_{\rm NL}$ using the integral of the likelihood with respect to the parameter.

Including the full covariance matrix, we find with the curvature method $f_{\rm NL}=-10^{+310}_{-270}$ at the $2\sigma$ level using frequentist confidence intervals, and $f_{\rm NL}=-10^{+270}_{-260}$ using Bayesian credible regions. Similarly for the wavelet method, $f_{\rm NL}=0^{+230}_{-260}$ with frequentist confidence intervals and $f_{\rm NL}=0^{+240}_{-240}$. Finally, we show that the two methods provide approximately uncorrelated estimates of $f_{\rm NL}$; this observation naturally suggests an improved estimator which combines the two methods. With this combined test, we find  $f_{NL}=-5\pm 85$ and $f_{NL}=-5\pm 175$ at the $1\sigma$ and $2\sigma$ level respectively.

\section*{Acknowledgements}

We are grateful to an anonymous referee for insisting on including the full covariance matrix. We acknowledge the use of the Legacy Archive for Microwave Background Data Analysis (LAMBDA). Support for LAMBDA is provided by the NASA Office of Space Science. Partial financial support from INAF (progetto di ricerca
``Non-Gaussian primordial perturbations: constraints from CMB and 
redshift surveys'') is acknowledged. FKH acknowledges financial support from the CMBNET Research Training Network. We acknowledge use of the HEALPix \cite{healpix} software and analysis package for deriving the results in this paper. This research used resources of the National Energy Research Scientific Computing Center, which is supported by the Office of Science of the U.S. Department of Energy under Contract No. DE-AC03-76SF00098.

\end{document}